\documentclass[prl, amsmath, amssymb, amsfonts, superscriptaddress, twocolumn, footinbib]{revtex4-1}

\usepackage{amsmath}
\usepackage{graphicx}
\usepackage{dcolumn}
\newcolumntype{d}[1]{D{.}{.}{#1}}

\usepackage{bm}
\usepackage{epstopdf}
\usepackage{graphicx}
\usepackage{stmaryrd}
\usepackage{float}
\usepackage{hyperref}
\usepackage{tabularx}
\usepackage{subfigure}
\usepackage[normalem]{ulem}
\usepackage[usenames, dvipsnames]{color}

\newcommand{\appropto}{\mathrel{\vcenter{
			\offinterlineskip\halign{\hfil$##$\cr
				\propto\cr\noalign{\kern2pt}\sim\cr\noalign{\kern-2pt}}}}}

\usepackage[normalem]{ulem}

\newcommand{\om}{\iffalse}

\newcommand{\ba}{\arraycolsep 0.3ex \begin{array}{rl}}
	\newcommand{\ea}{\end{array}}
\newcommand{\bc}{\begin{cases}}
	\newcommand{\ec}{\end{cases}}
\usepackage{color}
\newcolumntype{C}[1]{>{\centering\arraybackslash}p{#1}}

\begin{document}
	\title{Generating a topological anomalous Hall effect in a non-magnetic conductor}
	
	\author{James H. Cullen}
	\affiliation{School of Physics, The University of New South Wales, Sydney 2052, Australia}
		\affiliation{Australian Research Council Centre of Excellence in Low-Energy Electronics Technologies, The University of New South Wales, Sydney 2052, Australia}	
	\author{Pankaj Bhalla}
	\affiliation{Beijing Computational Science Research Center, Beijing 100193, China}
		\affiliation{Australian Research Council Centre of Excellence in Low-Energy Electronics Technologies, The University of New South Wales, Sydney 2052, Australia}	
	\author{E. Marcellina}
	\altaffiliation{Present address: School of Physical and Mathematical Sciences, Nanyang Technological University, 21 Nanyang Link, Singapore 637371. Email: emarcellina@ntu.edu.sg}
	\affiliation{School of Physics, The University of New South Wales, Sydney 2052, Australia}
		\affiliation{Australian Research Council Centre of Excellence in Low-Energy Electronics Technologies, The University of New South Wales, Sydney 2052, Australia}
	\author{A.~R.~Hamilton}
	\affiliation{School of Physics, The University of New South Wales, Sydney 2052, Australia}
		\affiliation{Australian Research Council Centre of Excellence in Low-Energy Electronics Technologies, The University of New South Wales, Sydney 2052, Australia}	
	\author{Dimitrie Culcer}
	\affiliation{School of Physics, The University of New South Wales, Sydney 2052, Australia}
		\affiliation{Australian Research Council Centre of Excellence in Low-Energy Electronics Technologies, The University of New South Wales, Sydney 2052, Australia}

\date{\today}

\begin{abstract}
The ordinary Hall effect is driven by the Lorentz force, while its anomalous counterpart occurs in ferromagnets. Here we show that the Berry curvature monopole of non-magnetic 2D spin-3/2 holes leads to a novel Hall effect linear in an applied \textit{in-plane magnetic field} $B_\parallel$. There is no Lorentz force hence no ordinary Hall effect, while all disorder contributions vanish to leading order in $B_\parallel$. This intrinsic phenomenon, which we term the \textit{anomalous planar Hall effect} (APHE), provides a non-quantized footprint of topological transport directly accessible in $p$-type semiconductors.
\end{abstract}

\maketitle


\textit{Introduction}. The ordinary Hall effect occurs in a classical magnetic field due to the Lorentz force and has been well understood since its discovery. In contrast, the understanding of the anomalous Hall effect (AHE) has been a key challenge for modern physics \cite{Nagaosa-AHE-2010}. Originally observed in ferromagnets, it was shown to exist in paramagnetic materials as well \cite{Culcer_AHE_PRB03}. The origins, relative importance and even existence of the manifold contributions to the AHE have spawned a debate spanning three quarters of a century \cite{Luttinger_AHE_PR58, Smit_SS_58, Berger_SJ_PRB1970, NozLew_SSSJ_JP73, Bruno, Dugaev_AHE_FM_Localization_PRB01, Nagaosa_japan, Culcer_AHE_PRB03, Haldane, Sinitsyn, AHE_vertex_PRL_2006, Nagaosa_AHE_PRB08, Sinitsyn_AHE_Review_JPCM08, Kovalev_Multiband_AHE_PRB09, Nagaosa-AHE-2010, Yang_SJ_SctUniversality_PRB10, Nomura_3DTI_QAHE_PRL11}. Fittingly, controversy surrounds it even as it opens new avenues of research \cite{Culcer_TI_AHE_PRB11, Ado, Ado_long, Liu2018, Keser_PRL2019, Yang_PRL2019}.

The band-structure or \textit{intrinsic} contribution to the AHE, nowadays understood to stem from the Berry curvature and have a topological origin, is quantized under certain circumstances\cite{Nagaosa_japan, Culcer_AHE_PRB03, Haldane, Sinitsyn, Yu_TIF_QAHE_Science2010, experimental_AHE, QAHE_exp2, precise_AHE, HighT_AHE, QAHE_review}. This places it in the expanding category of quantized topological responses, which include the quantum Hall and quantum spin-Hall effects \cite{Klitzling, Bernevig, QSH-HgTe_2007, Culcer_2DM_Review_2020}. Notably, quantized responses have only been observed in systems that are effectively one-dimensional, being intimately connected with the existence of edge states \cite{QSH-HgTe_2007, experimental_AHE, QAHE_exp2, precise_AHE, Collins2018}. Whereas the intrinsic AHE can be quantized in a two-dimensional semiconductor, such quantization is impossible to observe in practice, since (i) the dispersion involves an even number of Zeeman-split Kramers pairs that make opposite topological contributions, yielding a non-universal conductivity that is often density-dependent \cite{Jungwirth, Nunner, Sinitsyn, Nagaosa-AHE-2010} and (ii) disorder corrections are very strong \cite{AHE_vertex_PRL_2006, Culcer_InterbandCoh_PRB2017}. Nevertheless, we show here that, with appropriate design, the Berry curvature can be unambiguously identified in 2D transport even when the response is not quantized. 

\begin{figure}[tbp]
	{\includegraphics[width = \columnwidth]{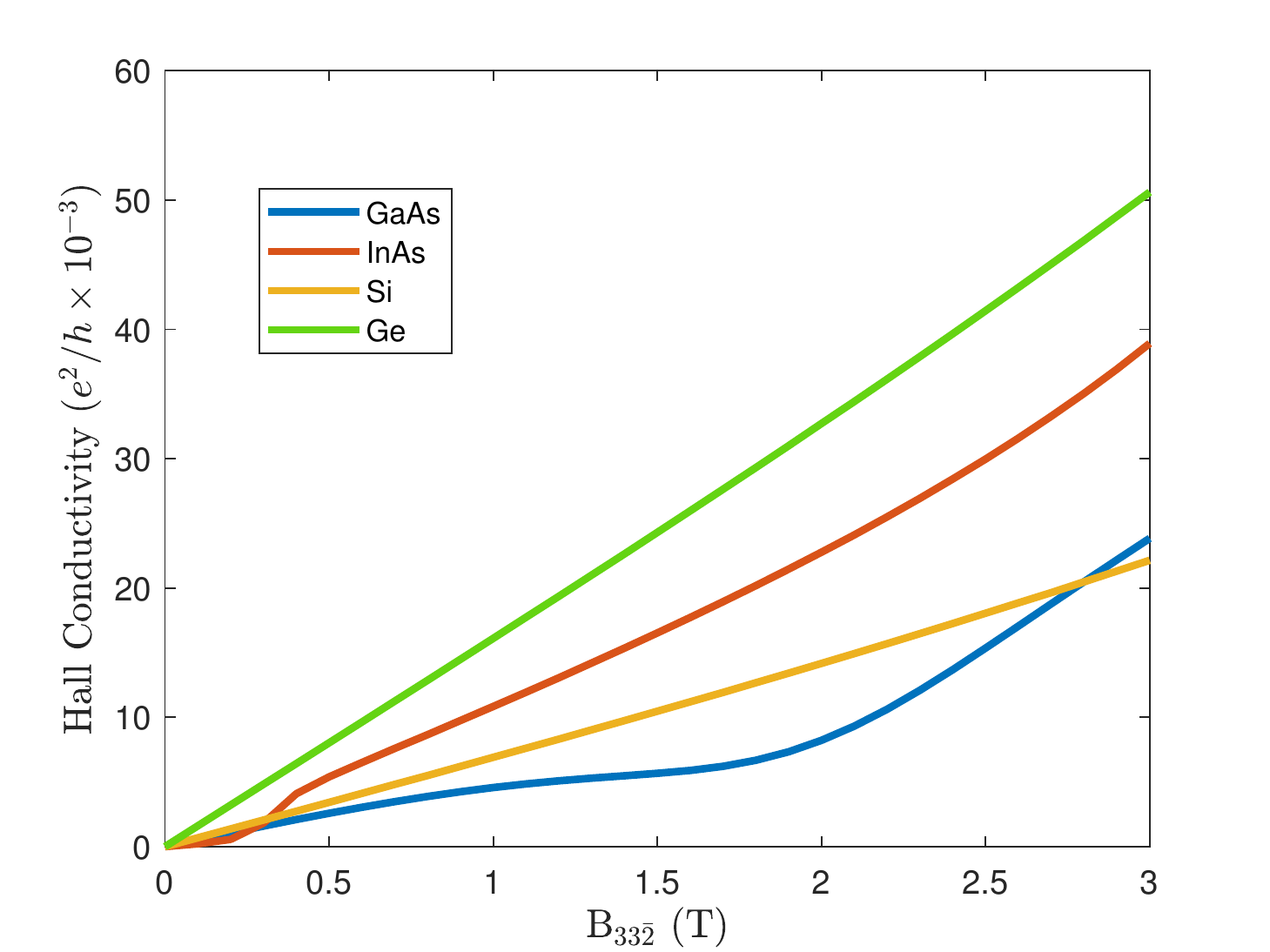}}
	\caption{\label{fig: 2Dplot1} Intrinsic Hall conductivity versus in-plane magnetic field for different materials, for a symmetric quantum well grown along (113) and well width of 20 nm. The Fermi energy is 5 meV, the carrier densities $\approx 5-25\times10^{10}$ cm$^{-2}$.
	}
\end{figure}

\begin{figure}[tbp]
	{\includegraphics[width = \columnwidth]{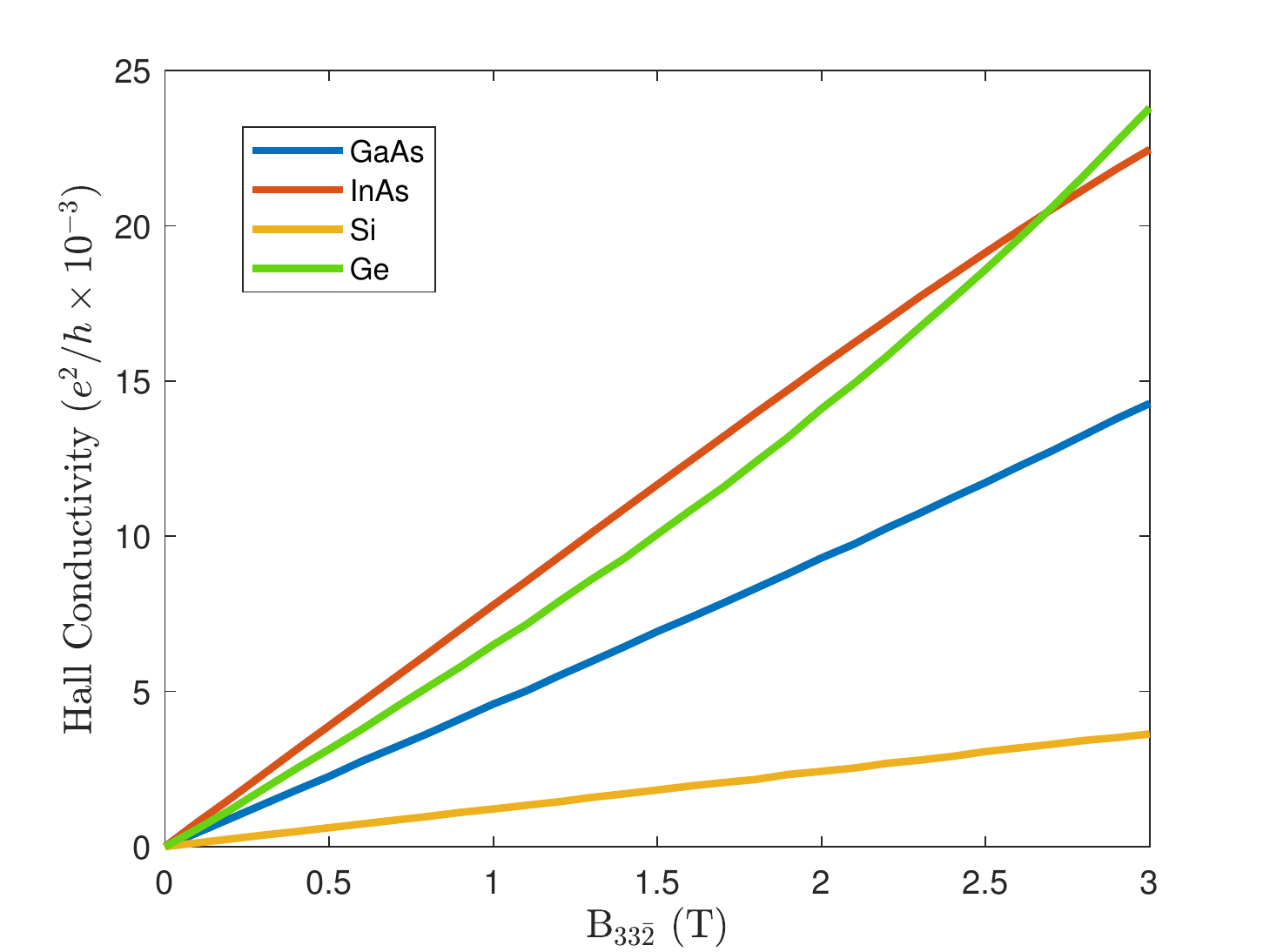}}
	\caption{\label{fig: 2Dplot2} Intrinsic Hall conductivity versus in-plane magnetic field for different materials for an asymmetric well grown along (113) with top gate field $E_z=5$ MV/m and well width of 20 nm. The Fermi energy is 5 meV, the carrier densities $\approx 5-25\times10^{10}$ cm$^{-2}$.
	}
\end{figure}

We show that a Hall effect can be generated in a two-dimensional heavy-hole system grown on a low-symmetry substrate when an \textit{in-plane} magnetic field $B_\parallel$ is applied. In the absence of an out-of-plane magnetic field and therefore of a Lorentz force, there is no ordinary Hall effect. Instead, an anomalous Hall effect occurs due to the finite Berry curvature of the spin-3/2 hole system, with no counterpart in spin-1/2 electron systems. The gap in the dispersion that enables the Hall response is opened by a shear Zeeman term in the Lande $g$-tensor \cite{PhysRevLett.113.236401, Tommy_gzx_PRB2016}. Our central result is the Hall conductivity $\sigma_{xy}$ linear in $B_\parallel$, shown in Figs.~\ref{fig: 2Dplot1} and \ref{fig: 2Dplot2} and expected from the Onsager relations. We refer to this phenomenon as the \textit{anomalous planar Hall effect} (APHE). Its origins are intrinsic and topological: we show that neither scalar nor extrinsic spin-orbit scattering contribute to leading order in $B_\parallel$. The APHE is a manifestation of relativistic quantum mechanics in classical transport mediated by topological terms in the crystal band structure, which reflect inter-band coherence induced by external electromagnetic fields. Since the equilibrium system is not magnetized the effect is tunable \textit{in situ} by altering the magnetic field orientation. It is observable in state-of-the-art hole samples, which are developing at a brisk pace, motivated in part by quantum computing applications \cite{Nichele_PRL2014, Nichele-Ensslin-2014-PRB, Brauns-Zwanenburg-2016-PRB, Matthias-Zwanenburg-2016-APL, Daisy-2016-NL, Akhgar-2016-NL, Fanming-Kouwenhoven-2016-NL, Nichele-Kouwenhoven-2017-PRL, Srinivasan-2017-PRL, Conesa-Boj-2017-NL, Li_APL2017, Li_NL2018, Vukusic_NL2018, Liles2018, Hendrickx2018, Silvano_PRL2018, Hendrickx2020, Tarucha_Holes2020}, where holes are actively investigated \cite{Kloeffel2013, Shim_NC2014, Jo-2017-PRB}, by their large spin-orbit coupling \cite{Moriya-2014-PRL, Nichele_PRL2017, Beukman_PRB2017}, interesting transport properties \cite{Liu_Hole_FQHE_PRL2014, HongMarcellina2018, Marcellina2018, Marcellina_PRB2020, Auerbach_2020}, and by unconventional properties stemming from their spin-3/2 nature \cite{Winkler2008, Chao-Xing-2008-PRB, Chesi-2011-PRL, Scholz_PRB2013, PhysRevB.89.161307, Tutul-2014-AP, Mawrie2014, Cuan-2015-EPL, Fu-2015-PRB, Biswas-2015-EP, Paul-2016-PRB, Mawrie_JPCM2017, Liang2017, Bladwell2019}.

We consider a two-dimensional hole gas grown along (113). Our findings hold for a multitude of low-symmetry growth directions, and (113) is chosen as an example. The ground state is represented by the heavy hole manifold, where heavy holes have spin projection $\pm 3/2$ onto the quantization axis, which is perpendicular to the plane. At normal transport densities the light hole manifold is not occupied. Our results are obtained using the Luttinger Hamiltonian rotated along the growth direction (113) with the vertical confinement modeled by the Bastard wave-function, given in the Supplement. Nevertheless, the underlying physics can be understood from the effective Hamiltonian for the lowest heavy hole sub-band $H_{hh} = \varepsilon_{0{\bm k}} + H_s + H_c$, where $\varepsilon_{0{\bm k}} = \hbar^2k^2/(2m^*)$, with $m^*$ the heavy-hole in-plane effective mass and ${\bm k}$ the in-plane wave vector. The term $H_s = (\hbar/2) {\bm \sigma}\cdot{\bm \Omega}_{\bm k}$, with ${\bm \sigma}$ the vector of Pauli spin matrices, captures the leading contributions to the Rashba and Zeeman effects, which stem from the spherical terms in the Luttinger Hamiltonian \cite{Winkler-2000-PRB, Shayegan-2002-PRB, Roland, Durnev_PRB2014, Elizabeth-2017-PRB, Miserev2017, Miserev2017a}, as well as the out-of-plane Zeeman term, which stems from the cubic symmetry of diamond and zincblende lattices \cite{Tommy_gzx_PRB2016}. These can be incorporated into an effective field ${\bm \Omega}_{\bm k}$. We assume $\Omega_F \tau \gg 1$, with $\Omega_F$ the value of ${\bm \Omega}_{\bm k}$ on the Fermi surface and $\tau$ the momentum relaxation time. Written in full,
\begin{equation}\label{Hs}
\arraycolsep 0.3ex
\begin{array}{rl}
\displaystyle H_s = & \displaystyle i\alpha(\sigma_+k_-^3 - \sigma_-k_+^3) + \Delta_1 (\sigma_+B_-k_-^2 + \sigma_-B_+k_+^2) \\ [1ex]
+ & \displaystyle \Delta_2 (\sigma_+B_+k_-^4 + \sigma_-B_-k_+^4) + \Delta_3 k^2 (\sigma_+B_+ + \sigma_-B_-) \\ [1ex]
+ & \displaystyle \Delta_4 (\sigma_+B_-k_+^2 + \sigma_-B_+k_-^2) + \Delta_{zx} \sigma_z B_x.
\end{array}
\end{equation}
with $\alpha$ is the Rashba spin-orbit constant in the spherical approximation, $k_\pm = k_x \pm i k_y$, $B_\pm = B_x \pm i B_y$, and $\Delta_1, \Delta_2, \Delta_3, \Delta_4$ are effective Lande $g$-factors, with $\Delta_3, \Delta_4$ only present in asymmetric wells \cite{Miserev2017, Miserev2017a}.
The shear Zeeman term $\Delta_{zx} \sigma_z B_x$ is vital \cite{PhysRevLett.113.236401, Tommy_gzx_PRB2016}, yielding an out-of-plane Zeeman splitting in response to an in-plane magnetic field that, henceforth, we assume to be $\parallel \hat{\bm x}$, where $\hat{\bm x}\parallel (33\bar{2})$. The eigenvalues of $H_0 + H_s$ are $\varepsilon_{{\bm k}\pm} = \varepsilon_{0{\bm k}} \pm \hbar |\Omega_{\bm k}|$. Cubic symmetry leads to the following additional spin-orbit terms
 \begin{equation}\label{Hextra}
	\arraycolsep 0.3ex
	\begin{array}{rl}
	\displaystyle H_c = & \eta_1 \, \sigma_z k_y + \eta_2 \, \sigma_z k_y(11 k_y^2 - 49 k_x^2) \\ [1ex]
	+ & \displaystyle i  \eta_3 \, (\sigma_+ k_- - \sigma_- k_+)
	\end{array}
 \end{equation}
Here $\eta_1$ has contributions from the Dresselhaus terms $\propto C_D, B_{D1}$ in Ref.~\onlinecite{Elizabeth-2017-PRB}, $\eta_2 \propto B_{D1}$, while $\eta_3$ and $\eta_4$ have contributions from both Dresselhaus and cubic-symmetry terms. We recall that Dresselhaus terms are absent in diamond lattices such as Si and Ge, but are noticeable in zincblende materials such as GaAs and InAs. 

We focus first on the Hall current due to $H_s$, which provides the dominant contribution. The effect of the much smaller terms contained in $H_c$ \cite{Elizabeth-2017-PRB}, as well as disorder, is discussed in the closing section. The Zeeman term for heavy holes also includes terms $\propto B_\parallel^3$, yet these are three orders of magnitude smaller than the $B_{\parallel}-$linear terms above, only becoming important for $B_{\parallel} \ge 30$ T. The coefficients in Eqs.~\ref{Hs}, \ref{Hextra} are functions of $k$ and decrease strongly at larger wave vectors. Such momentum-dependent Zeeman terms with different winding numbers are likewise specific to heavy holes, since these correspond to the $\pm 3/2$ projection of the hole spin-$3/2$ onto the quantization axis \cite{Marcellina_PRB2020}. They have no counterpart in electron systems. In our evaluations this $k$-dependence is circumvented by using the full Luttinger Hamiltonian.

\begin{figure}[tbp]
	\centering
	\begin{subfigure}[]
		\centering
		\includegraphics[width = 0.48\linewidth]{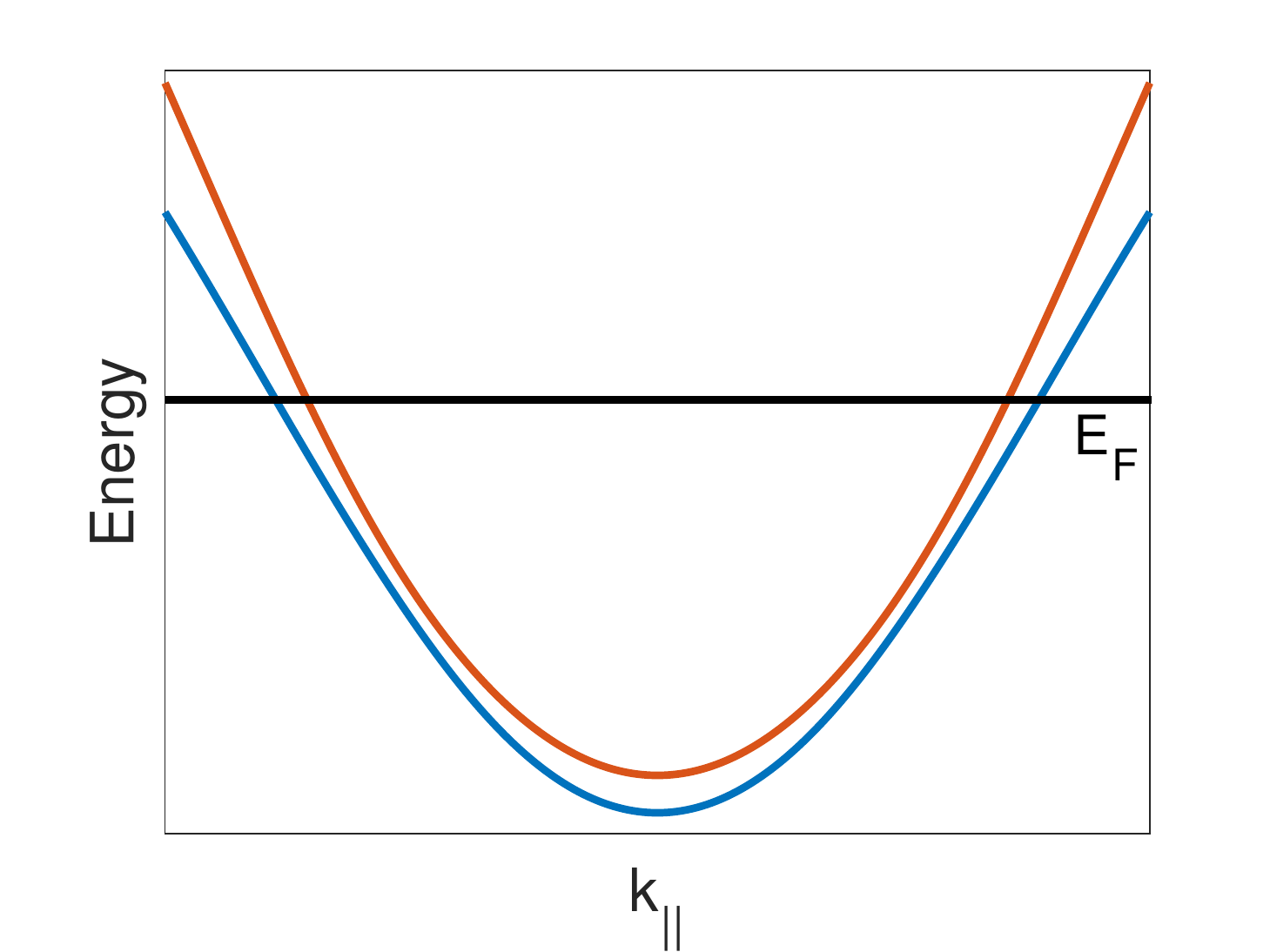}
		\label{Ez=0Disp}
	\end{subfigure}
	\begin{subfigure}[]
		\centering
		\includegraphics[width = 0.48\linewidth]{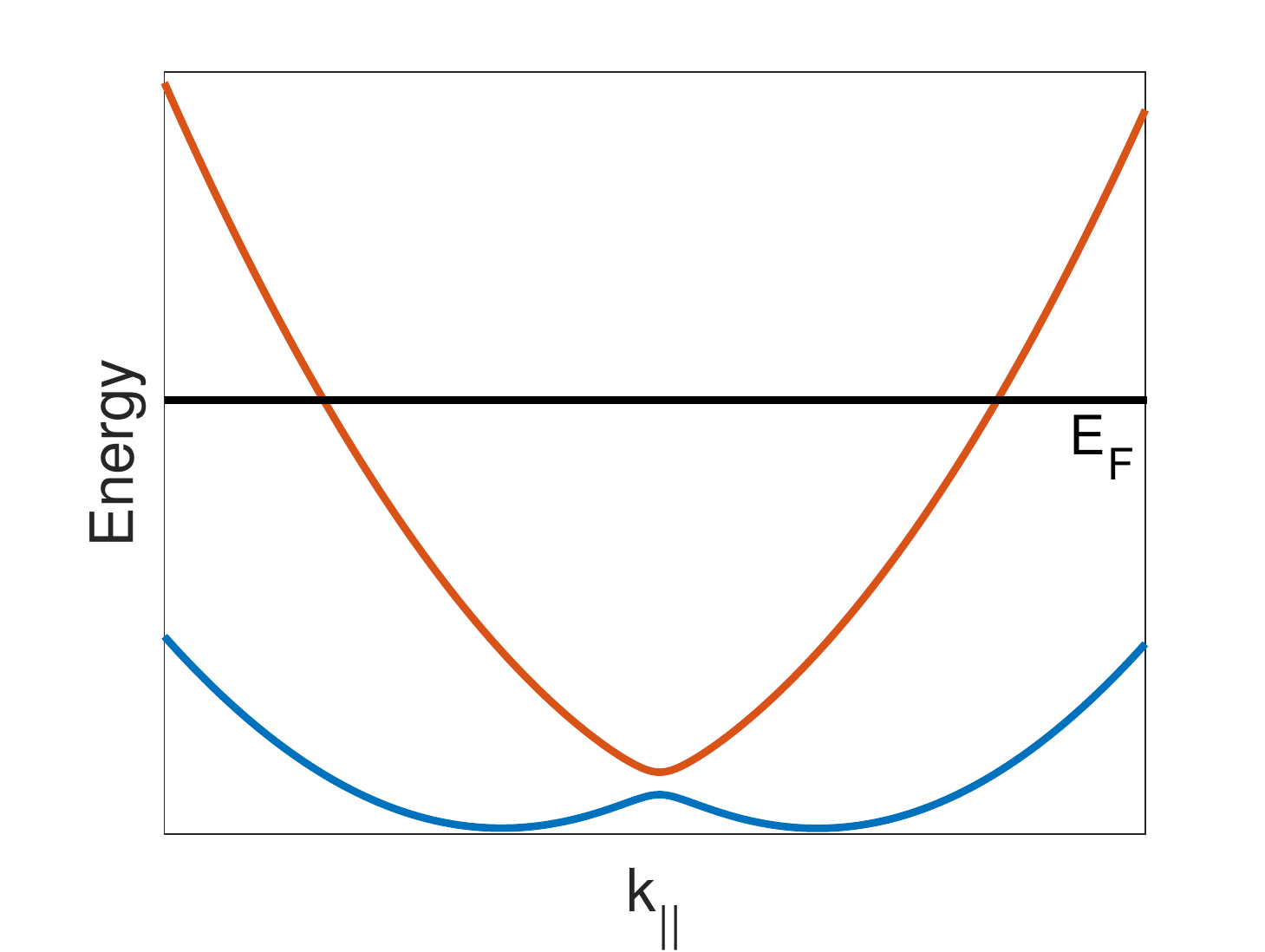}
		\label{Ez=/=0Disp}
	\end{subfigure}
	\caption{\label{Fig:Disp} The in-plane dispersion of the two spin-split HH subbands for a 2DHG grown along (113) with an in-plane magnetic field. (a) Symmetric well. (b) Asymmetric well.}
\end{figure}

Unlike the ordinary Hall effect driven by the Lorentz force, the charge dynamics here originate in the spin locking to the wave vector. As a hole is accelerated longitudinally its wave vector changes and it experiences a different spin-orbit field. As a result of this the spin undergoes a rotation, which in turn produces a change in the wave vector that is perpendicular to the longitudinal motion, in other words, a Hall current. The intrinsic Hall conductivity, derived below, is given by $\sigma_{xy}^{0} = - \frac{e^2}{h} \int d^2k/(2\pi) \, {\mathcal F}_z^n$, where the Berry curvature of sub-band $n$ is $\bm{\mathcal F}_n = -{\rm Im} \, \langle \partial u_n/\partial {\bf k}| \times |\partial u_n/\partial {\bf k}\rangle $, with $u_n$ the lattice-periodic Bloch wave function, whose role is played here by the envelope function. The full effect as a function of magnetic field, including its material dependence, is shown in Fig.~\ref{fig: 2Dplot1} for a symmetric well and in Fig.~\ref{fig: 2Dplot2} for a strongly asymmetric well, in which the Rashba interaction dominates. The form and behaviour of the Hall conductivity are understood by noting that (i) the shear Zeeman term opens a gap, which makes the Berry curvature nonzero; (ii) the spin-split bands have different Fermi wave vectors, and (iii) the sign of the Berry curvature is determined by the winding direction of the spin-orbit field ${\bm \Omega}_{\bm k}$, with the Rashba, in-plane Zeeman and the Dresselhaus terms all yielding the same sign. However, for different low symmetry growth directions these contributions can yield different signs. We first provide a \textit{pedagogical analytical explanation} for these two limiting cases: a symmetric well with a single in-plane Zeeman term expected to dominate at small densities, and a strongly asymmetric well. For simplicity we use constant coefficients to derive approximate analytical expressions for $\sigma_{xy}$, noting that this pedagogical approach is restricted to very small densities and magnetic fields. The dispersions for these two cases are sketched in Fig.~\ref{Fig:Disp}.

In a symmetric well in Si and Ge, given that there is no a-priori spin-orbit coupling, the Berry curvature is initially zero. When the in-plane magnetic field is applied, the Zeeman terms $\propto \Delta_1, \Delta_2$, give rise to a non-zero Berry curvature at each ${\bm k}$, while the Zeeman term $\propto \Delta_{zx}$ opens a gap in the spectrum. Interestingly the Berry curvature itself is independent of $B_\parallel$, although it depends explicitly on the in-plane and out-of-plane $g$-factors. Since the heavy-hole sub-band is now spin-split by the out-of-plane Zeeman interaction there are two different Fermi wave vectors. The difference between them is linear in the magnetic field at low fields, as shown in the Supplement, hence the conductivity is linear in $B_\parallel$. With only the $\Delta_1$ and $\Delta_{zx}$ terms, the Hall conductivity reads
\begin{equation}
\sigma_{xy} = \frac{e^2}{h}B_{||} \left[\frac{8\Delta_{zx}\Delta_1^2m^{*3}\epsilon_{F}\sqrt{4\Delta_1^2\epsilon_{F}^2+\frac{\hbar^4\Delta_{zx}^2}{m^{*2}}}}{\hbar^6(\Delta_{zx}^2+\frac{4\Delta_1^2m^{*2}\epsilon_F^2}{\hbar^4})^{\frac{3}{2}}}\right],
\end{equation}
which increases monotonically with $B_\parallel$. In GaAs and InAs, on the other hand, the Dresselhaus terms cause $\sigma_{xy}$ to be non linear at small fields but their contribution to the Berry curvature is eventually overwhelmed by the in-plane Zeeman terms as $B_\parallel$ increases. 

In an asymmetric well the Berry curvature is likewise zero in the absence of the out-of-plane Zeeman interaction. Nevertheless, the heavy hole sub-band is already spin-split by the strong Rashba spin-orbit interaction even before the magnetic field is turned on, so that there is a sizable difference between the two Fermi wave vectors. The Rashba interaction overwhelms all other terms, as shown recently \cite{Elizabeth-2017-PRB}, hence the plots for the asymmetric well increase monotonically for all materials. When the Zeeman term $\propto \Delta_{zx}$ opens a gap a significant Hall current emerges. With only the Rashba and $\Delta_{zx}$ terms, the Hall conductivity takes the form
\begin{equation}
\sigma_{xy} = \frac{e^2}{h}B_{||} \bigg(\frac{3\Delta_{zx}}{2\alpha}\bigg) \left(\frac{1}{k_{F+}^3}-\frac{1}{k_{F-}^3}\right).
\end{equation}
The Fermi wave vectors $ k_{F\pm}\approx (\frac{2m^*\epsilon_F}{\hbar^2}\pm\frac{4\sqrt[6]{54} \alpha m^{*\frac{5}{2}}\epsilon_F^{\frac{3}{2}}}{\hbar^5})^{1/2} $ differ due to the Rashba interaction, while their magnetic field dependence is negligible. Because of this, when the spin-orbit energy $\Omega_F > \Delta_{zx} B_x$ but is still much less than the kinetic energy, $\sigma_{xy}$ is approximately independent of spin-orbit strength. This insight is more general than just the Rashba case, and explains the relative smallness of the effect and its comparable size in all materials studied. The APHE is driven by the cubic-symmetry and bulk Zeeman terms, which are strongest in InAs and Ge. In a realistic sample we expect $\rho_{xy} \sim 100 - 500 \mu\Omega$.

\begin{figure}[tbp]
	\centering
	\includegraphics[width=0.85\linewidth]{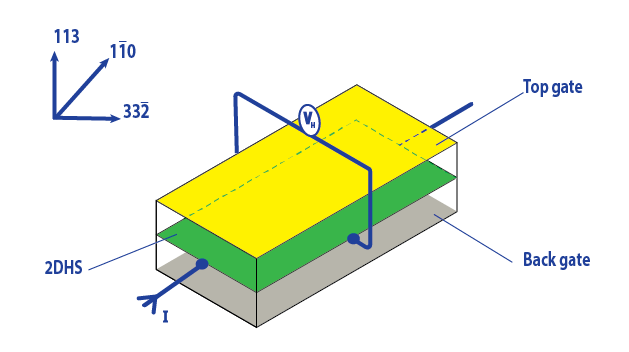}
	\caption{\label{Fig:Schem} Experimental setup for measuring the APHE. A magnetic field $\parallel (33\bar{2})$ will cause a Hall current. Rotating it to $(1\bar{1}0)$ will cause the Hall current to disappear.}
\end{figure}

\textit{Experimental measurement}. A schematic of an experimental setup that could be used to measure the APHE is shown in Fig.~\ref{Fig:Schem}. To identify the topological terms experimentally, one can start with a (113) QW with a top and bottom gate that enables full control of the Rashba interaction \cite{Papadakis-1999-Science,Li2016,PhysRevLett.113.236401}, and apply an in-plane magnetic field to introduce the Zeeman interactions \ref{Hs}. The topological terms can be detected through the Hall voltage, since it depends directly on $\sigma_{xy}$. As the magnetic field is rotated in the plane the Hall current will disappear, since $\Delta_{zy} = 0$, enabling one to turn the APHE on and off in situ. As the field is rotated out of the plane the APHE will give way to the ordinary Hall effect.


\textit{Methodology}. The full Hamiltonian $H = \varepsilon_{0{\bm k}} + H_s + H_E + U$. The driving electric field ${\bm E} \parallel \hat{\bm y}$ is contained in $H_E = e{\bm E}\cdot\hat{\bm r}$. The scattering potential $U({\bm r}) = \sum_I \bar{U} ({\bm r} - {\bm R}_I),$ includes both scalar and extrinsic spin-orbit scattering \cite{Bi2013}, where ${\bm R}_I$ indexes the random locations of impurities, and the scattering potential due to a single impurity is denoted by $\bar{U}({\bm r})$. In Fourier space, 
$ \bar{U}_{{\bm k}{\bm k}'} = \mathcal{U}_{{\bm k}{\bm k}'} \openone + \mathcal{V}_{{\bm k}{\bm k}'}$,
where $\mathcal{U}_{{\bm k}{\bm k}'}$ represents a matrix element between plane waves, and, $\mathcal{V}_{{\bm k}{\bm k}'} = -\frac{i\lambda}{2} \, {\bm \sigma} \cdot ({\bm \omega}_{\bm k}
\times{\bm k}' - {\bm \omega}_{{\bm k}'} \times{\bm k}) \, \mathcal{U}_{{\bm k}{\bm k}'}$
where ${\bm \omega}_{{\bm k}} = k^3 \, (\cos 3 \theta, \sin 3
\theta, 0)$, assuming $\lambda k_F^4 \ll 1$. Since the Rashba Hamiltonian is derived from the Luttinger Hamiltonian in the spherical approximation it has the same form along (113) as along (001), and thus the extrinsic spin-orbit scattering term has the same form as in Ref.~\onlinecite{Bi2013}. As written in the Pauli basis, the spin dependent term $\mathcal{V}_{{\bm k}{\bm k}'}$ points out of the plane. We note that $\lambda$ for holes has not been determined quantitatively. We consider short-range scattering off uncorrelated impurities, with the average of $\langle {\bm k}n|\hat{U}|{\bm k}'n'\rangle\langle{\bm k}'n'\hat{U}|{\bm k}n\rangle$ over impurity configurations $n_i |\bar{U}_{{\bm  k}{\bm k}'}^{nn'}|^2/V$, where $n_i$ is the impurity density and $V$ the crystal volume. 

We derive a quantum kinetic equation as described in Refs.~\onlinecite{Culcer_InterbandCoh_PRB2017, Vasko}. The  density matrix $\rho$ is found in the basis $\{{\bm k}n\}$, where $n$ represents the band index. To determine the charge current we require $f_{\bm k}$, the part of the density matrix diagonal in wave vector, since the current operator is diagonal in ${\bm k}$. From the quantum Liouville equation, $\partial\rho/\partial t + (i/\hbar) \, [H, \rho] = 0$, we obtain the following kinetic equation describing the time evolution of $f_{\bm k}$:
\begin{equation}\label{Kinetic-e} 
\frac{\partial f_{\bm k}}{\partial t}+\frac{i}{\hbar}\left[H_{0{\bm k}} + H_{\text{Z}}, f_{\bm k}\right]+\hat{J}(f_{\bm k})=\mathcal{D}_{E,{\bm k}},
\end{equation}
where the scattering term in the Born approximation
\begin{equation} \label{J0}
\hat{J}(f_{\bm k})\!=\!\left\langle\frac{1}{\hbar^2}\int^{\infty}_0\!dt'[\hat{U},e^{-\frac{iH_0t'}{\hbar}}[\hat{U},\hat{f}(t)]e^{\frac{iH_0t'}{\hbar}}]\right\rangle_{{\bm k}{\bm k}},
\end{equation}  
and the driving term $\mathcal{D}_{E,{\bm k}} =  \frac{D f_{0{\bm k}}} {D {\bm k}}$. The covariant derivative $ \frac{D f_{0{\bm k}}} {D {\bm k}} = \frac{\partial f_{0{\bm k}}} {\partial {\bm k}} - i \, [ \mathcal{\bm R}_{\bm k}, f_{0{\bm k}}] $ arises from the ${\bm k}$-dependence of the basis functions. The Berry connection matrix elements $\mathcal{\bm R}_{\bm k}^{mm'} = \langle u_{\bm k}^{m}\vert i \frac{\partial u_{\bm k}^{m'}}{\partial {\bm k}}\rangle$, with $m \ne m'$ necessarily, are interband matrix elements of the position operator, which also appear in the current density operator ${\bm j} = -(e/\hbar) \, D H_{0{\bm k}}/D{\bm k}$. In an external electric field one may decompose $f_{\bm k} = f_{0{\bm k}} + f_{E{\bm k}}$, where $f_{0{\bm k}}$ is the equilibrium density matrix and $f_{E{\bm k}}$ is a correction to first order in the electric field. The equilibrium density matrix is $f_{0{\bm k}} = (1/2) \, [(f_{{\bm k}+} + f_{{\bm k}-}) \openone + \tilde{\sigma}_z (f_{{\bm k}+} - f_{{\bm k}-})]$, where $f_{{\bm k}\pm}$ represent the Fermi-Dirac distributions over subband energies $\varepsilon_{{\bm k}\pm}$, and the tilde in $\tilde{\sigma}_z$ denotes the basis of eigenstates of the band Hamiltonian. In linear response one may replace $f_{{\bm k}} \rightarrow f_{0{\bm k}}$ in the driving term $\mathcal{D}_{E,{\bm k}}$. With $f_{0{\bm k}}$ known and $\mathcal{D}_{E,{\bm k}}$ on the right-hand side of Eq.~(\ref{Kinetic-e}), we obtain $f_{E{\bm k}}$. By taking the trace with the current operator, the longitudinal and transverse components of the current are found. The off-diagonal part of the density matrix contains the intrinsic term $S_{E{\bm k}}^{mm'} = - i\hbar \mathcal{P} [D^{mm'}_{E{\bm k}}(\varepsilon^m_{\bm k} - \varepsilon^{m'}_{\bm k})]$, with $\mathcal{P}$ the principal part, and this yields directly the intrinsic Hall conductivity as introduced above. 

\textit{Disorder contributions}. Disorder is responsible for a complex series of contributions to the Hall conductivity, which in related models tend to reduce the Hall effect or cancel it altogether \cite{AHE_vertex_PRL_2006, Culcer_InterbandCoh_PRB2017}. Disorder contributions fall into two categories: those stemming from scalar disorder and from spin-dependent disorder. The former exist because of inter-band coherence induced by the electric field, which causes even scalar disorder to contribute to the Hall effect through an anomalous driving term
\begin{align}\label{Dprime}
D'^{\, mm''}_{E{\bm k}} = & \displaystyle  \frac{\pi n_i}{\hbar} 
\sum_{m'{\bm k}'} \mathcal{U}^{mm'}_{{\bm k}{\bm k}'} \mathcal{U}^{m'm''}_{{\bm k}'{\bm k}} \left[ (
n_{E{\bm k}}^{m} -  n_{E{\bm k'}}^{m'} ) \delta(\varepsilon^m_{\bm k} - \varepsilon^{m'}_{{\bm k}'})\right. \nonumber\\
&\left. + \displaystyle (n_{E{\bm k}}^{m''} - n_{E{\bm k}}^{m'}) 
\delta(\varepsilon^{m''}_{{\bm k}} - \varepsilon^{m'}_{{\bm k}'}) \right].
\end{align}
Here we have singled out the band-diagonal term in the density matrix, $n_{E{\bm k}}^{mm'} \propto \delta^{mm'}$. To find $D'^{\, mm''}_{E{\bm k}}$, we first solve for $n_{E{\bm k}}$, feed it into the scattering term Eq.~\ref{J0}, and take the band off-diagonal part. When calculated explicitly this contribution is identically zero for both symmetric and strongly asymmetric wells for both the spherical Rashba interaction and the Zeeman terms.

The spin-dependent disorder term $\mathcal{V}_{{\bm k}{\bm k}'}$ gives rise to four different contributions: skew scattering, side-jump, anomalous spin precession and the anomalous scattering term. These were described in detail in Ref.~\onlinecite{Bi2013} for the spin-Hall effect, and the method we follow here is exactly analogous. Due to the large winding numbers involved the anomalous spin precession and the anomalous scattering terms are zero for holes, while skew scattering and side-jump contribute only to order $B_\parallel^3$. We also find the contribution from $J_\lambda(S_{E\parallel})$ to be imaginary. Hence, with the Rashba interaction evaluated in the spherical approximation, and in the absence of Dresselhaus terms there are no disorder contributions to the APHE. 

We investigate next the additional contributions to the Hall current due to the Dresselhaus and cubic-symmetry Rashba term. The term $\propto \eta_3$ in Eq.~\ref{Hextra} is linear in wave vector, therefore their contribution to the intrinsic part of the Hall current is cancelled exactly by the scalar disorder term in the same way as the customary linear Rashba term in electron systems \cite{AHE_vertex_PRL_2006, Culcer_InterbandCoh_PRB2017}. 
It does not contribute to the extrinsic signal because they generate a spin-orbit field that is purely in the plane. The Dresselhaus terms make a contribution to the Berry curvature, which is noticeable only in a symmetric wells, as discussed above. Since the Dresselhaus spin-orbit field $\propto \eta_1, \eta_2$ points entirely out of the plane, it does not contribute to the extrinsic signal through the anomalous scattering term $J_{\Omega\lambda}(n_{E\parallel})$. It does not contribute to the Hall current through the anomalous driving term either. The only way a nonzero contribution to the current can emerge is through products of the Dresselhaus terms and cubic symmetry terms, which will then be multiplied by the small parameter $\lambda k_F^4 \ll 1$. Therefore, in Si and Ge the extrinsic contribution to the APHE vanishes altogether, while in InAs and GaAs it is negligibly small.

In summary, we have shown that low-symmetry growth hole nanostructures exhibit a purely Berry-curvature driven Hall effect in response to an in-plane magnetic field. The APHE may open new pathways in the electrical operation of spin qubits and spin-orbit torques.

\acknowledgments

This work was funded by the Australian Research Council Centre of Excellence in Future Low-Energy Electronics Technologies. PB acknowledges the Chinese Postdocs Science Foundation grant No. 2019M650461 and NSAF China grant No. U1930402 for financial support. We thank Assa Auerbach, Abhisek Samanta, Daniel Arovas, John Schliemann and Michael Fuhrer for helpful discussions.

\appendix

\begin{widetext}

\section*{APHE analytical solutions for the 2x2 HH Hamiltonian}
For this approach the Zeeman terms and the Rashba SOC coefficient are all approximated as constants.
\subsection*{APHE with Rashba}

For the first case we set $g_{zx}\neq0$, $g_1=0$, $g_2=0$, $\alpha\neq0$. $\alpha$ is the Rashba spin-orbit coupling constant, $g_1$ and $g_2$ are the two in-plane Zeeman terms.\\
\\
For this case the 2x2 heavy hole Hamiltonian is\\
\begin{equation}
H=\frac{\hbar^2 k^2}{2m^{*}}+i\alpha(k^3_-\sigma_+-k^3_+\sigma_-)+g_{zx}B_x\sigma_z\hspace{3mm},
\end{equation}
here $\sigma_{\pm}$ are the raising/lowering operators and $\sigma_z$ is the diagonal Pauli matrix for puesdospin-$\frac{1}{2}$ space for HH states (spin $\pm\frac{3}{2}$). \\

\noindent The eigenstate energies of this system are\\
\begin{equation}
\epsilon=\frac{\hbar^2 k^2}{2m^{*}}\pm\sqrt{g_{zx}^2B_x^2+\alpha^2k^6}\hspace{3mm}.
\end{equation}
The z component of the berry curvature for each eigenstate is
\begin{equation}
\Omega_{z\pm}=\mp\frac{9g_{zx}\alpha^2B_xk^4}{2(g_{zx}^2B_x^2+\alpha^2k^6)^{\frac{3}{2}}}\hspace{3mm}.
\end{equation}
The intrinsic contribution to the AHE is defined as\\
\begin{equation}
\sigma_{i j}=-\varepsilon_{i j \ell} \frac{e^{2}}{\hbar} \sum_{n} \int \frac{d \boldsymbol{k}}{(2 \pi)^{d}} f(\varepsilon_{n}(\boldsymbol{k})) \Omega_{n}^{\ell}(\boldsymbol{k})\hspace{3mm},
\end{equation}
where $\hat{\varepsilon}$ is the asymmetric tensor and $f$ is the Fermi-Dirac distribution. Assuming the system is at a very low temperature, this integral will give the following AHE conductivity
\begin{equation}
\sigma_{xy}=\frac{3e^2}{2h}g_{zx}B_x\left(\frac{1}{\sqrt{g_{zx}^2B_x^2+\alpha^2k_{F+}^6}}-\frac{1}{\sqrt{g_{zx}^2B_x^2+\alpha^2k_{F-}^6}}\right)\hspace{3mm}.
\end{equation}
Where $k_{F\pm}$ are the Fermi wave vectors of the spin split sub-bands. To find $k_{F\pm}$, (2) must be solved for k.
The analytical solution to this is very complex but it can be approximated as\\
\begin{equation}
k_{F\pm}\approx \left(\frac{2m^*\epsilon_F}{\hbar^2}\pm\frac{4\sqrt[6]{54} \alpha m^{*\frac{5}{2}}\epsilon_F^{\frac{3}{2}}}{\hbar^5}\right)^{1/2}\hspace{3mm}.
\end{equation}
\\
Analysis of the exact solution shows that $k_F$ has very little dependence on the magnetic field and can be assumed to be constant for realistic magnetic fields.\\
\\
Hence for small magnetic fields  $g_{zx}B_x\ll \beta k_F^3$, the intrinsic AHE conductivity can be approximated as\\
\begin{equation}
\sigma_{xy}\approx\frac{3e^2}{2h}\left(\frac{B_xg_{zx}}{\alpha}\left(\frac{1}{k_{F+}^3}-\frac{1}{k_{F-}^3}\right)+\frac{B_x^3g_{zx}^3}{\alpha^2}\left(\frac{1}{k_{F+}^9}-\frac{1}{k_{F-}^9}\right)\right)\hspace{3mm}.
\end{equation}
\\
The $B_x^3$ term is generally many orders of magnitude smaller than the linear term. So for small magnetic fields $\sigma_{xy}\propto B_x$.\\
\subsection*{APHE with in-plane Zeeman}
For this case we set $g_{zx}\neq0$, $g_1\neq0$, $g_2=0$, $\alpha=0$ and so the 2x2 heavy hole Hamiltonian is\\
\begin{equation}
H=\frac{\hbar^2 k^2}{2m^{*}}+g_1(B_+k^2_+\sigma_-+B_-k^2_-\sigma_+)+g_{zx}B_x\sigma_z\hspace{3mm},
\end{equation}
where $B_{\pm}=B_x\pm iB_y$. For this calculation we only consider magnetic fields in the x direction.\\

\noindent The eigenstate energies of this system are\\
\begin{equation}
\epsilon=\frac{\hbar^2 k^2}{2m^{*}}\pm B_x\sqrt{g_{zx}^2+g_1^2k^4}\hspace{3mm}.
\end{equation}{}
The z component of the berry curvature for each eigenstate is
\begin{equation}
\Omega_{\pm}=\mp\frac{2g_{zx}g_1^2k^2}{(g_{zx}^2+g_1^2k^4)^{\frac{3}{2}}}\hspace{3mm}.
\end{equation}
Using the low temperature approximation, the intrinsic AHE conductivity for this system is
\begin{equation}
\sigma_{xy}=\frac{e^2}{h}g_{zx}\left(\frac{1}{\sqrt{g_{zx}^2+g_1^2k_{F+}^4}}-\frac{1}{\sqrt{g_{zx}^2+g_1^2k_{F-}^4}}\right)\hspace{3mm}.
\end{equation}
For this system, $k_F$ will have a strong dependency on $B_x$. Solving (9) gives the following solution for $k_F$
\begin{equation}
(k_{F\pm})^2=\frac{4\hbar^2m^*\epsilon_{F}\mp4m^{*2} B_x\sqrt{4g_1^2\epsilon_{F}^2+\frac{\hbar^4g_{zx}^2}{m^{*2}}-4g_{zx}^2g_1^2B_x^2}}{2\hbar^4-8m^{*2}g_1^2B_x^2}\hspace{3mm}.
\end{equation}
Here we approximate the solution by assuming $B_x$ is small so we only take the linear magnetic field terms into account.
\begin{equation}
(k_{F\pm})^2\approx\frac{2\hbar^2m^*\epsilon_{F}\mp2m^{*2} B_x\sqrt{4g_1^2\epsilon_{F}^2+\frac{\hbar^4g_{zx}^2}{m^{*2}}}}{\hbar^4}\hspace{3mm}.
\end{equation}
Using this solution the approximate term for the intrinsic AHE conductivity is
\begin{equation}
\sigma_{xy}\approx\frac{e^2}{h}B_x\left(\frac{8g_{zx}g_1^2m^{*3}\epsilon_{F}\sqrt{4g_1^2\epsilon_{F}^2+\frac{\hbar^4g_{zx}^2}{m^{*2}}}}{\hbar^6(g_{zx}^2+\frac{4g_1^2m^{*2}\epsilon_F^2}{\hbar^4})^{\frac{3}{2}}}\right)\hspace{3mm}.
\end{equation}
(14) shows that the intrinsic contribution to the Hall Conductivity is linear in $B_x$. This approximation is valid up to $\approx2$T.\\
\\
\subsubsection*{Second in-plane Zeeman term}
We also considered the rare case where the second in-plane Zeeman term $g_2$ dominates. For this calculation we set $g_{zx}\neq0$, $g_1=0$, $g_2\neq0$, $\alpha=0$ and so the 2x2 heavy hole Hamiltonian is\\
\begin{equation}
H=\frac{\hbar^2 k^2}{2m^{*}}+g_2(B_-k^4_+\sigma_-+B_+k^4_-\sigma_+)+g_{zx}B_x\sigma_z\hspace{3mm}.
\end{equation}
The eigenstate energies of this system are\\
\begin{equation}
\epsilon=\frac{\hbar^2 k^2}{2m^{*}}\pm B_x\sqrt{g_{zx}^2+g_2^2k^8}\hspace{3mm}.
\end{equation}{}
The z component of the berry curvature for each eigenstate is
\begin{equation}
\Omega_{\pm}=\mp\frac{8g_{zx}g_2^2k^6}{(g_{zx}^2+g_2^2k^8)^{\frac{3}{2}}}\hspace{3mm}.
\end{equation}
In the low temperature approximation, the intrinsic AHE conductivity for this system is
\begin{equation}
\sigma_{xy}=2\frac{e^2}{h}g_{zx}\left(\frac{1}{\sqrt{g_{zx}^2+g_2^2k_{F+}^8}}-\frac{1}{\sqrt{g_{zx}^2+g_2^2k_{F-}^8}}\right)\hspace{3mm}.
\end{equation}
Solving (16) for $k$ to find $k_F$ gives a non-trivial solution, but for small magnetic fields it can be approximated as
\begin{equation}
(k_{F\pm})^2\approx\frac{2m^*\epsilon_F}{\hbar^2}\mp\frac{8m^{*3}g_2B_x}{\hbar^6}(5g_{zx}^2B_x^2+\epsilon_F^2)\hspace{3mm}.
\end{equation}
Using this solution, the intrinsic contribution to the Hall conductivity can be approximated as
\begin{equation}
\sigma_{xy}\approx\frac{e^2}{h}B_x\left(\frac{512m^{*6}\epsilon_F^5g_2^3g_{zx}}{\hbar^{12}(g_{zx}^2+g_2^2\frac{16m^{*4}\epsilon_F^4}{\hbar^8})^{\frac{3}{2}}} \right)\hspace{3mm}.
\end{equation}
This approximation is valid for fields up to $\approx2$T. So again, the AHE conductivity is linear in $B_x$.

\section*{Numerical density matrix calculations}
To calculate the size of the APHE, the intrinsic AHE conductivity was numerically calculated using the 4x4 Luttinger Hamiltonian rotated to (113). Only the 4x4 Hamiltonian was used, as all other bands are separated by large energy gaps and have little influence on the behaviour of heavy holes.\\

\noindent We only considered the ground state of the QW. This does limit the accuracy of the calculation, but for the purpose of demonstrating the existence and rough size of the APHE we decided that this was enough.\\

\subsection*{Rotated 4x4 Luttinger}
The 4x4 Luttinger Hamiltonian used in this calculation was found using the following approximation\\
\begin{equation}
H=\mu\bigg[\gamma_1k^2 - 2\gamma_2\bigg((J_x^2-\frac{1}{3}J^2)k_x^2+\text{cp}\bigg)-4\gamma_3\bigg(\{J_x,J_y\}k_xk_y+\text{cp}\bigg) \bigg]\hspace{3mm}.
\end{equation}
Where $\gamma_1,\,\gamma_2\,\&\,\gamma_3$ are the Luttinger parameters, $\mu$ is the mass term $\frac{\hbar^2}{2m_0}$ and the anticommutator is defined as $\{A,B\}=\frac{1}{2}(AB+BA)$. After rotating the co-ordinates, equation (21) gave following Hamiltonian
\begin{equation}
H = 
\begin{bmatrix}
P+Q & 0 & L+\sqrt{3}\kappa\mu_BB_- & M \\
0 & P+Q & M* & -L*+\sqrt{3}\kappa\mu_BB_+ \\
L*+\sqrt{3}\kappa\mu_BB_+ & M & P-Q & 2\kappa\mu_BB_- \\
M* & -L+\sqrt{3}\kappa\mu_BB_- & 2\kappa\mu_BB_+ & P-Q 
\end{bmatrix}
\end{equation}
where
\begin{equation*}
\begin{aligned}
P=\gamma _1 \mu  \left(k_x^2+k_y^2+k_z^2\right) \hspace{3mm},
\end{aligned}
\end{equation*}

\begin{equation*}
\begin{aligned}
Q=\frac{\mu}{121} \left(8 \gamma _2 \left(18 \sqrt{2} k_x k_z+5 k_x^2+11 k_y^2-16 k_z^2\right)+3 \gamma _3 \left(-48 \sqrt{2} k_x k_z+27 k_x^2+11 k_y^2-38 k_z^2\right)\right)\hspace{1mm},
\end{aligned}
\end{equation*}  

\begin{equation*}
\begin{aligned}
L=\frac{\sqrt{3} \,\mu}{121} \bigg(\gamma _2 \left(66 i \sqrt{2} k_x k_y-108 k_x k_z-15 \sqrt{2} k_x^2+44 i k_y k_z-33 \sqrt{2} k_y^2+48 \sqrt{2} k_z^2\right)\\
+\gamma _3  \left(-66 i \sqrt{2} k_x
k_y-134 k_x k_z+15 \sqrt{2} k_x^2+198 i k_y k_z+33 \sqrt{2} k_y^2-48 \sqrt{2} k_z^2\right)\bigg)\hspace{1mm},
\end{aligned}
\end{equation*}

\begin{equation*}
\begin{aligned}
M=\frac{\sqrt{3} \, \mu}{121} \bigg(\gamma _2 \left(18 k_x \left(\sqrt{2} k_z+11 i k_y\right)+5 k_x^2+66 i \sqrt{2} k_y k_z+11 k_y^2-16 k_z^2\right)\\
-2 \gamma _3 \left(-22 i k_x k_y+9 \sqrt{2} k_x k_z+63
k_x^2+33 i \sqrt{2} k_y k_z-55 k_y^2-8 k_z^2\right)\bigg)\hspace{1mm},
\end{aligned}
\end{equation*}

\noindent$k_x$ and $k_y$ are the in-plane wave vectors and $k_z$ is out-of-plane wave vector. The Hamiltonian will look different depending on the choice of $k_x$ and $k_y$, but the physics will remain unchanged. For this Hamiltonian $\vec{x} \, || \, (33\bar{2})$, $\vec{y} \, || \, (\bar{1}10)$ and $\vec{z} \, || \, (113)$.

\subsection*{Dresselhaus terms}
Because Dresselhaus SOC is significant in both GaAs and InAs, we included it in the numerical calculation for accuracy. Some preliminary analytical calculations were done that showed Dresselhaus contributions can give a non-linear result for the APHE at low magnetic field strengths as seen in Fig 1. These are not included here as they are non-trivial and don't add much value.\\
\\
For the numerical calculation we included the following Dresselhaus terms 
\begin{equation}
H_D=-\frac{2}{\sqrt{3}}C_D[k_x\{J_x,J_y^2-J_z^2\}+\text{cp}]-B_{D1}[k_x(k_y^2-k_z^2)J_x+\text{cp}]\hspace{3mm}.
\end{equation}
These two terms in our rotated co-ordinates are shown below 
\begin{equation}
\begin{split}
-C_D
&\left[
\begin{matrix}
-\frac{6}{11} \sqrt{\frac{2}{11}} k_y & \frac{9}{22 \sqrt{11}} \left(4 i k_x+4 k_y+5 i \sqrt{2} k_z\right) \\
\frac{9}{22 \sqrt{11}} \left(-4 i k_x+4 k_y-5 i \sqrt{2} k_z\right) & \frac{6}{11} \sqrt{\frac{2}{11}} k_y \\
\frac{1}{22} \sqrt{\frac{3}{11}} \left(-21 i k_x-3 k_y+4 i \sqrt{2} k_z\right) & -\frac{1}{44} i \sqrt{\frac{3}{11}} \left(5 \sqrt{2} k_x-35 i \sqrt{2} k_y-48 k_z\right) \\
\frac{1}{44} \sqrt{\frac{3}{11}} \left(-5 i \sqrt{2} k_x+35 \sqrt{2} k_y+48 i k_z\right) & \frac{1}{22} i \sqrt{\frac{3}{11}} \left(21 k_x+3 i k_y-4 \sqrt{2} k_z\right) \\
\end{matrix}
\right.\\
\\
&\left.\begin{matrix}
\frac{1}{22} i \sqrt{\frac{3}{11}} \left(21 k_x+3 i k_y-4 \sqrt{2} k_z\right) & \frac{1}{44} \sqrt{\frac{3}{11}} \left(5 i \sqrt{2} k_x+35 \sqrt{2} k_y-48 i k_z\right) \\
\frac{1}{44} i \sqrt{\frac{3}{11}} \left(5 \sqrt{2} k_x+35 i \sqrt{2} k_y-48 k_z\right) & \frac{1}{22} \sqrt{\frac{3}{11}} \left(-21 i k_x-3 k_y+4 i \sqrt{2} k_z\right) \\
\frac{18}{11} \sqrt{\frac{2}{11}} k_y & \frac{3}{22 \sqrt{11}} \left(-21 i k_x+3 k_y+4 i \sqrt{2} k_z\right) \\
\frac{3}{22 \sqrt{11}} \left(21 i k_x+3 k_y-4 i \sqrt{2} k_z\right) & -\frac{18}{11} \sqrt{\frac{2}{11}} k_y \\
\end{matrix}\right]\hspace{3mm},
\end{split}
\end{equation}

\begin{equation}
-B_{D1}
\begin{bmatrix}
\frac{3}{44\sqrt{11}}A & 0 & \frac{\sqrt{3}}{44\sqrt{11}}B & 0\\
0 & -\frac{3}{44\sqrt{11}}A & 0 & \frac{\sqrt{3}}{44\sqrt{11}}B^*\\
\frac{\sqrt{3}}{44\sqrt{11}}B^* & 0 & \frac{1}{44\sqrt{11}}A & \frac{1}{22\sqrt{11}}B\\
0 & \frac{\sqrt{3}}{44\sqrt{11}}B & \frac{1}{22\sqrt{11}}B^* & -\frac{1}{44\sqrt{11}}A\\
\end{bmatrix}\hspace{3mm},
\end{equation}
where
\[
A=k_y\left(96 k_x k_z-49 \sqrt{2} k_x^2+\sqrt{2} \left(11 k_y^2+16 k_z^2\right)\right)\hspace{3mm},
\]

\[
\begin{aligned}
B=-k_x^2 \left(15 k_y+59 i \sqrt{2} k_z\right)+i k_x \left(10 i \sqrt{2} k_y k_z+33 k_y^2+12 k_z^2\right)-15 i k_x^3-84 k_y k_z^2\\
+11 i \sqrt{2} k_y^2 k_z+33 k_y^3+16 i \sqrt{2} k_z^3\hspace{1mm}.
\end{aligned}
\]
These Dresselhaus terms were ignored in our calculations with asymmetric QWs. This is because the size of their contributions are negligible when compared with Rashba SOC, and also because the $k_z^3$ terms are not Hermitian when calculated with the wave function used.
\subsection*{Numerical Intrinsic Calculation}
To define the system in the confinement direction we used the Bastard wave function. The following diagonal Hamiltonian was used for this calculation\\
\begin{equation}
H=\frac{\hbar^2k_z^2}{2m^*}+eFz \hspace{3mm},
\end{equation}
where $m^*$ is the out-of-plane effective mass and $F$ is the gate voltage. The bastard wave function for this Hamiltonian with an infinite QW is
\begin{equation}
\phi(z)=N \cos{\frac{\pi z}{L}}\exp{\left(-\beta\left(\frac{z}{L}+\frac{1}{2}\right)\right)},\hspace{3mm} \frac{|z|}{L}<\frac{1}{2}\hspace{3mm},
\end{equation}
where $\beta$ is a variational parameter. The $\vec{z}$ wave functions (27) for heavy and light holes are used to calculate the matrix terms in the 4x4 Hamiltonian.\\

\noindent This approach is not completely accurate as there are $k_z$ dependent terms in the off-diagonal of the Hamiltonian. However, these terms are small so they were treated as perturbations.\\
\\
The Berry curvature is defined as
\begin{equation}
\vec{\Omega}_n=-im\left\langle\frac{\partial u_n}{\partial \vec{k}}|\times|\frac{\partial u_n}{\partial \vec{k}}\right\rangle\hspace{3mm}.
\end{equation}{}
\noindent Finding analytical terms for the eigenstates of the Hamiltonian is very difficult. So the following equation was used to calculate the z component of the Berry curvature\\
\begin{equation}
\Omega^z_n= -\text{im}\left[\sum_{m} \frac{\langle u_n|\frac{d\hat{H}}{dk_x}| u_m\rangle \langle u_m |\frac{d\hat{H}}{dk_y}| u_n\rangle}{(\epsilon_n-\epsilon_m)^2}\right], \hspace{3mm} n\neq m \hspace{1mm}.
\end{equation}
$\frac{d\hat{H}}{dk}$ is trivial once the Hamiltonian is defined, so (29) is far less computationally intensive. (29) was used to calculate the Berry curvature of the 2 spin-split sub-bands, for an array of values of $k_{||}$. This was then integrated numerically to find the Hall conductivity.
\begin{equation}
\sigma_{i j}=-\varepsilon_{i j \ell} \frac{e^{2}}{\hbar} \sum_{n} \int \frac{d \boldsymbol{k}}{(2 \pi)^{d}} f(\varepsilon_{n}(\boldsymbol{k})) \Omega_{n}^{\ell}(\boldsymbol{k})\hspace{3mm}.
\end{equation}
Similar to the analytical calculation, the low temperature approximation was used for the Fermi distribution. The integration was done using the trapezoidal rule, because it is not computationally intensive and is sufficiently accurate if enough points in $k_{||}$ space are defined. This calculation was repeated for different magnetic fields and materials to make Figs 1 \& 2.\\

\end{widetext}

\end{document}